\documentclass[twocolumn]{aastex631}

\newcommand{\kms}{\rm km~s^{-1}}
\newcommand{\kmsmpc}{\rm km~s^{-1}~Mpc^{-1}}
\newcommand{\Msun}{{\rm M}_{\odot}}
\newcommand{\Mstar}{M_{*}}

\newcommand{\dn}{{\rm D}_{n}4000}

\begin{document}

\title{The Velocity Dispersion Function for Quiescent Galaxies in Massive Clusters from IllustrisTNG}

\correspondingauthor{Jubee Sohn}
\email{jubee.sohn@snu.ac.kr} 

\author[0000-0002-9254-144X]{Jubee Sohn}
\affiliation{Astronomy Program, Department of Physics and Astronomy, Seoul National University, 1 Gwanak-ro, Gwanak-gu, Seoul 08826, Republic of Korea} 
\affiliation{SNU Astronomy Research Center, Seoul National University, Seoul 08826, Republic of Korea} 

\author[0000-0002-9146-4876]{Margaret J. Geller}
\affiliation{Smithsonian Astrophysical Observatory, 60 Garden Street, Cambridge, MA 02138, USA} 

\author[0000-0002-1327-1921]{Josh Borrow}
\affiliation{Department of Physics and Astronomy, University of Pennsylvania, 209 South 33rd Street, Philadelphia, PA, USA, 19104}

\author[0000-0001-8593-7692]{Mark Vogelsberger}
\affiliation{Department of Physics and Kavli Institute for Astrophysics and Space Research, Massachusetts Institute of Technology, Cambridge, MA 02139, USA}

\begin{abstract}
We derive the central stellar velocity dispersion function for quiescent galaxies in 280 massive clusters with $\log (M_{200} / \Msun) > 14$ in IllustrisTNG300. The velocity dispersion function is an independent tracer of the dark matter mass distribution of subhalos in galaxy clusters. Based on the IllustrisTNG cluster catalog, we select quiescent member subhalos with a specific star formation rate $< 2 \times 10^{-11}$ yr${^-1}$ and stellar mass $\log (\Mstar / \Msun) > 9$. We then simulate fiber spectroscopy to measure the stellar velocity dispersion of the simulated galaxies; we compute the line-of-sight velocity dispersions of star particles within a cylindrical volume that penetrates the core of each subhalo. We construct the velocity dispersion functions for quiescent subhalos within $R_{200}$. The simulated cluster velocity dispersion function exceeds the simulated field velocity dispersion function for $\log \sigma_{*} > 2.2$, indicating the preferential formation of large velocity dispersion galaxies in dense environments. The excess is similar in simulations and in the observations. We also compare the simulated velocity dispersion function for the three most massive clusters with $\log (M_{200} / \Msun) > 15$ with the observed velocity dispersion function for the two most massive clusters in the local universe, Coma and A2029. Intriguingly, the simulated velocity dispersion functions are significantly lower for $\log \sigma_{*} > 2.0$. This discrepancy results from 1) a smaller number of subhalos with $\log (\Mstar / \Msun) > 10$ in TNG300 compared to the observed clusters, and 2) a significant offset between the observed and simulated $\Mstar - \sigma_{*}$ relations. The velocity dispersion function offers a unique window on galaxy and structure formation in simulations. 
\end{abstract}

\section{INTRODUCTION} \label{sec:intro}

The central velocity dispersion of quiescent galaxies has a long history as a fundamental observable \citep{Faber76, Djorgovski87, Wake12, Bogdan15, Schechter15}. The stellar mass velocity dispersion relation, a projection of the fundamental plane, is an important scaling relation \citep{Hyde09, Cappellari13, Belli14, Cappellari16, Zahid16, Cannarozzo20, Damjanov22}. The central stellar velocity dispersion is also a measure of the dark matter velocity dispersion and mass \citep{vanuitert13, Schechter15, Zahid16, Zahid18, Utsumi20, Seo20}.

In the crowded fields of clusters of galaxies, central velocity dispersions are relatively insensitive to the biases that may affect photometry and the consequent derivation of stellar masses for member galaxies \citep{Bernardi13, Bernardi17, Miller21, Li22}. Cluster luminosity and mass functions are both affected by these photometric biases. The velocity dispersion function for the quiescent population thus complements the more traditional luminosity and mass functions. 

Observational determinations of the field velocity dispersion function for quiescent galaxies include \citet{Sheth03, Choi07, Bernardi10, Montero17} and \citet{Sohn17b}. Velocity dispersion functions for observed clusters include \citet{Munari16, Sohn17a} and \citet{Sohn20a}. Differences between the field and cluster velocity dispersion functions reflect differences in the evolution of the population as a function of the environment. Clusters, especially the most massive systems, contain an excess of galaxies with large velocity dispersion \citep{Sohn17a, Sohn20a}. In fact, the central velocity dispersion of the Brightest Cluster Galaxy (BCG) is correlated with the cluster velocity dispersion \citep{Sohn20b, Sohn21}. This link between the velocity dispersion function and environment prompts our exploration of the cluster velocity dispersion function in the IllustrisTNG simulations for comparison with observations.

Observational determination of the velocity dispersion function has some limitations. One issue is the selection of the quiescent population (see e.g., \citealp{Williams09, Moresco13, Damjanov18, Damjanov19}). Spectroscopic (selection based on the indicator $\dn$) and photometric (based on UVJ color) approaches yield consistent velocity dispersion functions for the quiescent population \citep{Sheth03, Choi07, Sohn17b}. Another underlying issue is the removal of objects where systemic rotation may contribute to the central velocity dispersion (see e.g., \citealp{vanuitert13}). The complexity of the correction for sample incompleteness is a subtle issue in the observational determination of the velocity dispersion function that requires careful attention \citep{Sohn17b}. Sample selection is more straightforward in simulated samples like IllustrisTNG; it is possible to select a velocity dispersion-limited sample of objects directly rather than starting with a magnitude-limited set.

\citet{Sohn24} compute stellar velocity dispersions of quiescent galaxies in the IllustrisTNG simulations \citep{Pillepich18, Nelson19}. The simulated velocity dispersions are insensitive to the computational approach. In comparison with observations, there is a puzzling offset between the simulated and observed stellar mass - velocity dispersion scaling reaction in the sense that the simulations underestimate the velocity dispersion at fixed stellar mass by $\sim 0.3$ dex. The observed relation clearly provides a benchmark for testing the simulations. The slopes of the simulated and observed scaling relations are the same. As in the Illustris-1 \citep{Nelson15} explored by \citet{Zahid18}, the stellar velocity dispersion traces both the dark matter halo velocity dispersion and mass. These relations suggest that the central velocity dispersions provide a route toward a deeper understanding of galaxies and their dark matter halos.

As an additional test of IllustrisTNG, we explore the velocity dispersion function for simulated massive clusters. We explore only the zero redshift snapshot as a basis for direct comparison with existing observations. The simulations are not tuned {\it a priori} to match the central velocity dispersions of the quiescent population. Thus the comparison with both the observed scaling relation and the velocity dispersion functions are measures of the fidelity of the simulated quiescent population.

We describe the IllustrisTNG-300 simulation in Section \ref{sec:data}. In Section \ref{sec:result}, we derive the stellar velocity dispersion functions of massive clusters in the IllustrisTNG300 simulation. We compare the velocity dispersion functions in field and cluster environments within TNG 300 in Section \ref{sec:fld}. We also compare the observed and simulated velocity dispersion functions in Section \ref{sec:obs}. We conclude in Section \ref{sec:conclusion}. We use the IllustrisTNG run cosmology (consistent with \citealp{Planck16}), i.e., $H_{0} = 67.74~\kmsmpc$, $\Omega_{m} = 0.3089$, and $\Omega_{\Lambda} = 0.6911$ throughout.

\section{IllustrisTNG} \label{sec:data}

We use IllustrisTNG to derive simulated velocity dispersion functions. IllustrisTNG is a set of cosmological magnetohydrodynamic simulations for exploring the formation and evolution of galaxies and larger structures \citep{Marinacci18, Naiman18, Nelson18, Pillepich18, Springel18, Nelson19}. IllustrisTNG improves on Illustris \citep{Vogelsberger14} by implementing a refined feedback model. IllustrisTNG includes multiple simulations that cover three different volumes with three different mass resolutions. We choose IllustrisTNG-300, which covers the largest volume with a length of 300 $h^{-1}$ Mpc. We use the highest resolution simulation (TNG300-1) in IllustrisTNG-300, with a mass resolution of $m_{\rm baryon} = 11 \times 10^{6}~\Msun$ and $m_{\rm DM} = 59 \times 10^{6}~\Msun$. Hereafter, we refer to this simulation as TNG300. 

We use TNG300 because it contains a large number of massive galaxy cluster halos. IllustrisTNG provides a list of cluster/group halos identified with a standard friends-of-friends (FoF) algorithm \citep{Huchra82}. With a linking length $b = 0.2$\footnote{A fractional linking length with respect to the mean separation of particles}, the FoF algorithm is first applied to dark matter; other types of particles are assigned to the nearest cluster/group halos. In TNG300, there are 280 halos with $\log (M_{200} / \Msun) > 14$ including 3 halos with $\log (M_{200} / \Msun) > 15$. These massive cluster halos are unique in TNG300. They do not appear in the companion volumes for IllustrisTNG, which are smaller in box size. TNG300 is thus a testbed for comparing the simulations with the velocity dispersion functions derived from Coma and A2029, the most massive clusters in the local universe with $\log (M_{200} / \Msun) > 15$ \citep{Sohn17a}. 

The TNG300 catalogs also list the subhalos belonging to the cluster/group halos. These subhalos are identified by the \texttt{SUBFIND} algorithm \citep{Springel01}. We identify galaxy subhalos with $\log (M_{*} / \Msun) > 9$ within each cluster halo; there are 462 to 11748 member subhalos within each cluster halo. We impose this mass limit to correspond roughly to the limit of dense spectroscopic surveys (e.g., \citealp{Sohn17a, Sohn20b}). 

\begin{figure}
\centering
\includegraphics[scale=0.29]{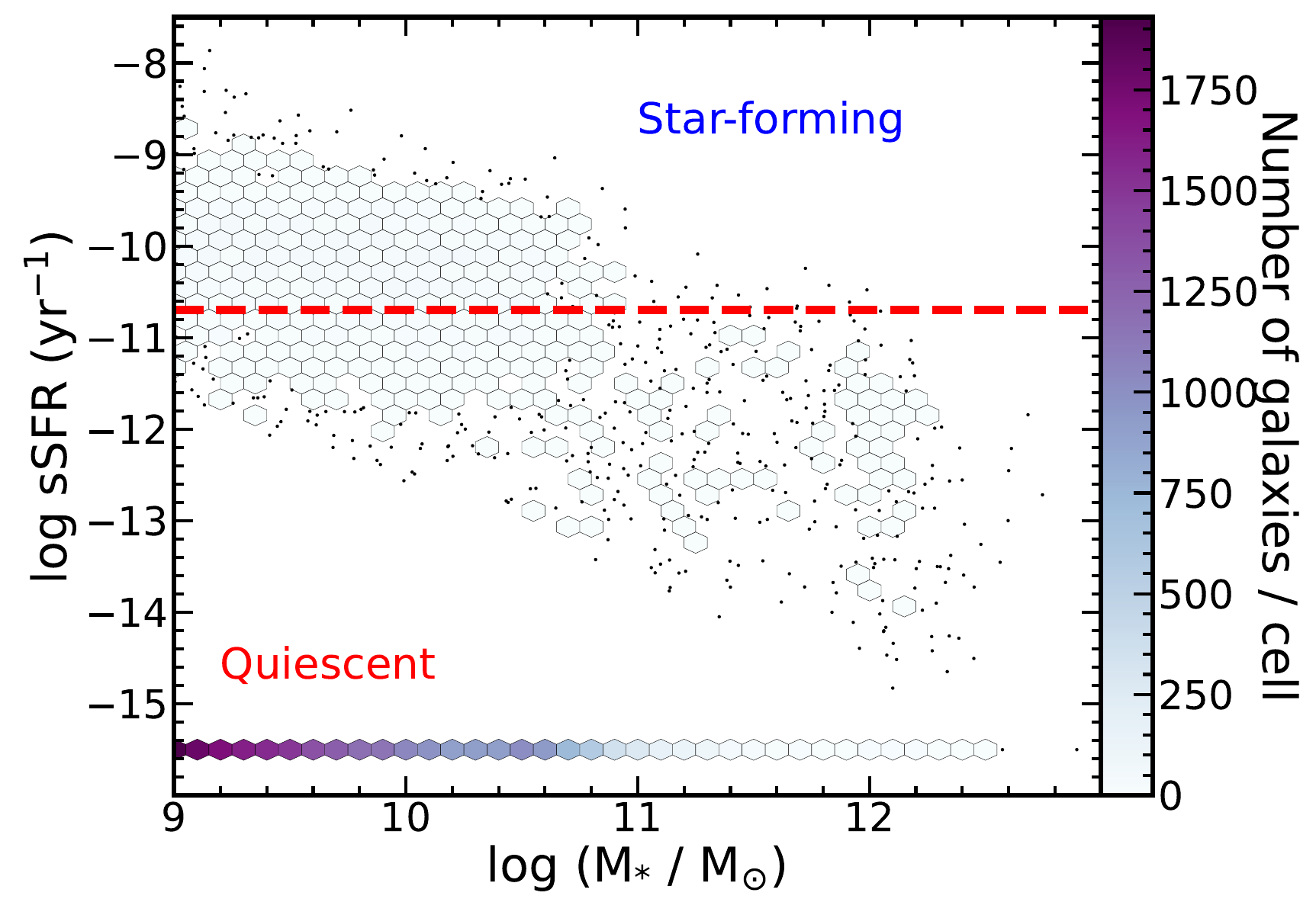}
\caption{Specific star formation rate as a function of stellar mass for all member subhalos within 280 massive clusters in IllustrisTNG-300. The horizontal line indicates the sSFR = $2 \times 10^{-11}$ yr$^{-1}$ we use to identify quiescent galaxies. }
\label{fig:ssfr}
\end{figure}

We select quiescent subhalos to investigate the velocity dispersion distribution. For star-forming subhalos, the disk rotation may affect measurements of the velocity dispersion. Figure \ref{fig:ssfr} displays the specific star formation rate as a function of stellar mass. There are subhalos without star formation (SFR $= 0$), and we allocate indicative sSFRs of $-15.0$ to mark them in the plot. A darker color indicates a higher number density. We define quiescent subhalos with the specific star formation rate smaller than $2 \times 10^{-11}$ yr$^{-1}$ following \citet{Sohn24}. 

The subhalos within massive cluster halos are mostly quiescent. The fraction of star-forming subhalos with sSFR $> 2 \times 10^{-11}$ yr$^{-1}$ varies between 0 to 13\%. Overall, only 3.5\% of cluster subhalos have significant star formation. Thus, the definition of the quiescent galaxy has little impact on the velocity dispersion distributions. In each cluster, there are 21 to 409 quiescent subhalos with $\log (M_{*} / \Msun) > 9$. 

\begin{figure}
\centering
\includegraphics[scale=0.31]{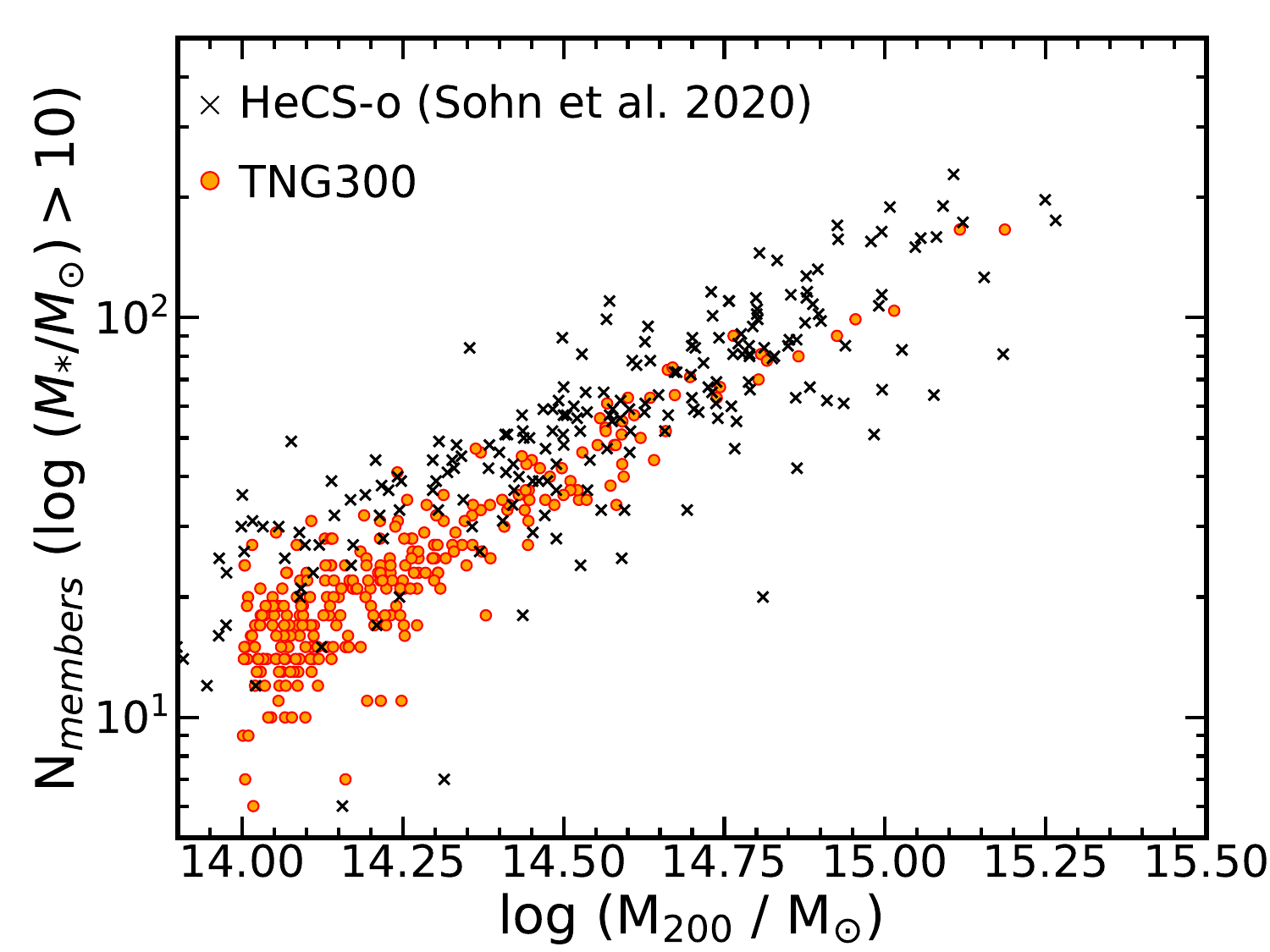}
\caption{Number of quiescent subhalos with stellar mass $\log (M_{*} / \Msun) > 10$ and within $R_{200}$ in the HeCS-omnibus sample (black crosses, \citealp{Sohn20b}) and in TNG300 (red circles). }
\label{fig:nsub}
\end{figure}

Figure \ref{fig:nsub} shows the number of member subhalos with $\log (M_{*} / \Msun) > 10$ as a function of cluster halo mass. Red circles show the 280 clusters in TNG300. Black crosses display the 225 clusters in the HeCS-omnibus cluster sample, a large spectroscopic compilation of clusters with $z < 0.3$ \citep{Sohn20a}. For both samples, we count the number of subhalos within $R_{200}$. We select quiescent galaxies with $\dn > 1.5$ from the HeCS-omnibus sample following \citet{Sohn24}. $\dn$ is a spectroscopic indicator of the age of stellar populationn(e.g., \citealp{Zahid17, Sohn17a, Sohn17b}). The spectroscopic surveys of HeCS-omnibus clusters are, of course, not 100\% complete. Thus, the numbers of member galaxies for HeCS-omnibus clusters in Figure \ref{fig:nsub} are lower limits. 

The number of subhalos in simulated clusters is significantly different from the number of observed cluster galaxies. A simulated cluster typically contains 8 - 40\% fewer quiescent subhalos with $\log (M_{*} / \Msun) > 10$. However, the mass function for subhalos with $\log (M_{*} / \Msun) > 9$ simulated clusters exceeds the abundance in observed clusters by 30 - 80\%. We revisit these issues further as they are reflected in the simulated and observed velocity dispersion functions (Section \ref{sec:obs}). 

We measure the stellar velocity dispersions of quiescent galaxy subhalos following \citet{Sohn24}. In \citet{Sohn24}, we compute the velocity dispersion based on various definitions; we vary the aperture, the viewing axes, and the measurement technique. This exploration demonstrates that systematic uncertainties are $\sim 10\%$. Here, we use the velocity dispersion based on member stellar particles identified by the \texttt{SUBFIND} algorithm within a 10 kpc aperture from the subhalo center along with $z-$axis, i.e., $\sigma_{*, 10~kpc, z, m-weight}$ in the notation of \citet{Sohn24}. We compute the mass-weighted velocity dispersion depending on the mass of each stellar particle. 

Previous observational studies often use velocity dispersions measured within a fiducial 3 kpc (e.g., \citealp{Zahid16, Sohn17a, Sohn17b, Sohn20b}), close to the fiber aperture. However, simulated measurements of velocity dispersions within this small aperture can be affected by gravitational softening. Thus, we use a larger aperture (i.e., 10 kpc) to measure the velocity dispersion. We note that the systematic difference between 3 kpc and 10 kpc velocity dispersions of TNG300 subhalos is less than 5\%. Hereafter, we refer to this $\sigma_{*, 10~kpc, z, m-weight}$ as $\sigma_{*}$. 

\begin{figure*}
\centering
\includegraphics[scale=0.48]{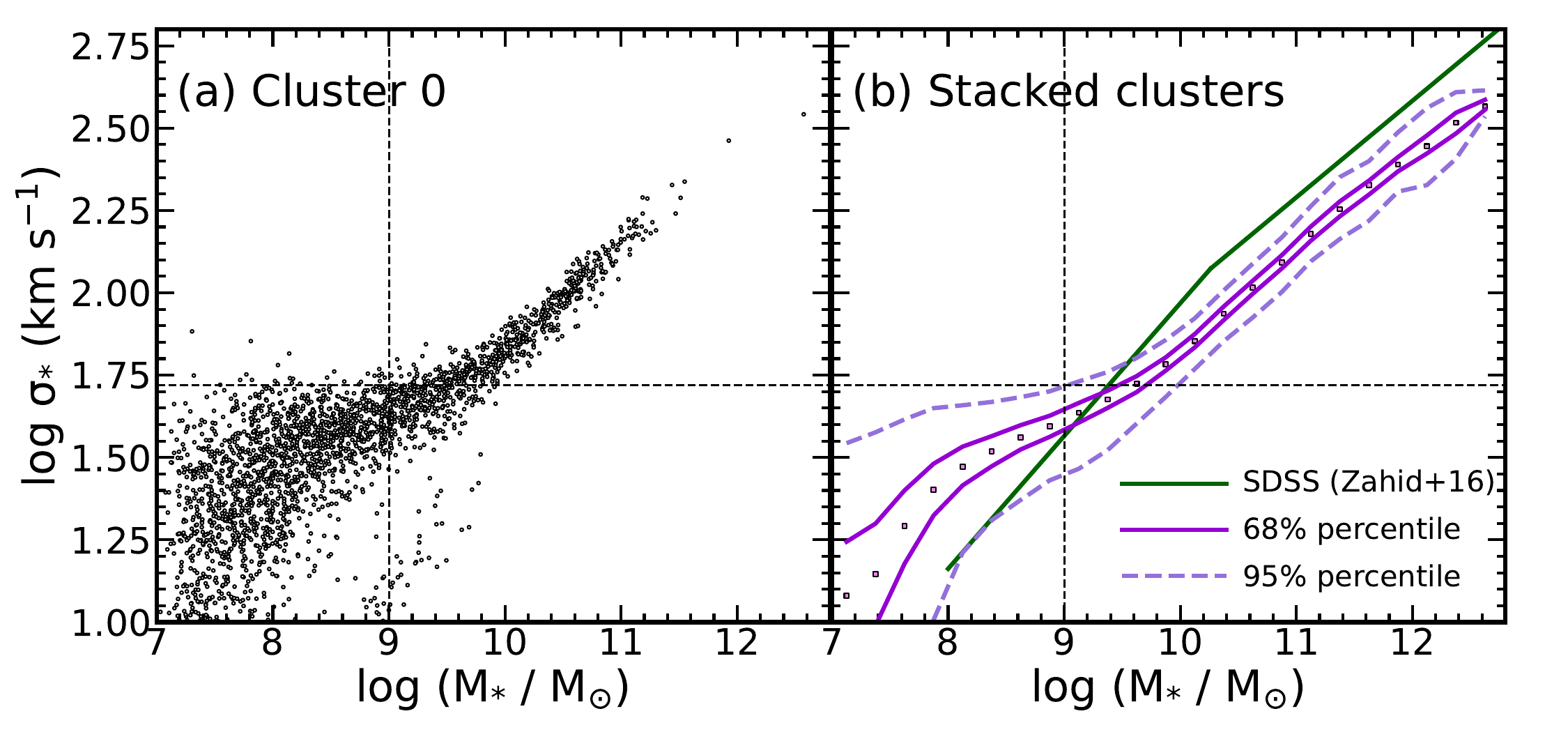}
\caption{(a) 1D stellar velocity dispersions within a 10 kpc aperture as a function of the stellar mass of subhalos within the most massive cluster halo in IllustrisTNG-300. (b) Same as (a), but based on all member subhalos in 280 cluster halos. Squares show the median distribution. Dotted and dashed lines denote the central 68\% and 95\% of the $\sigma_{*}$ distributions. The solid green line marks the same relation derived from SDSS quiescent galaxies \citep{Zahid16}. }
\label{fig:msigma}
\end{figure*}

Figure \ref{fig:msigma} (a) displays the $\sigma_{*}$ to stellar mass ($\Mstar$) relation for member subhalos in the most massive cluster halo in TNG300. Stellar velocity dispersions of subhalos generally increase in proportional to their stellar mass. The scatter in $\sigma_{*}$ is larger at $\log (\Mstar / \Msun) < 9$. 

Figure \ref{fig:msigma} (b) shows the similar $\sigma_{*}$ to $\Mstar$ relations based on all subhalos within the 280 cluster halos in TNG300. Squares display the median $\sigma_{*}$ distribution. Solid and dashed lines show the boundaries including central 68\% and 95\% of velocity dispersions. Only a very small fraction of subhalos lie outside the dashed lines. Because of the larger scatter in the $\sigma_{*}$ at low stellar mass, the sample we constructed is less complete. 

In Figure \ref{fig:msigma} (b), the solid green line displays the observed relation between $\sigma_{*}$ and $\Mstar$ from \citet{Zahid16}. Based on SDSS quiescent galaxies with $\dn > 1.5$ in the redshift range $z < 0.2$, \citet{Zahid16} derived the best-fit broken power laws. The quiescent galaxies in observations generally exhibit larger velocity dispersions compared to simulated subhalos at a given stellar mass. \citet{Sohn24} demonstrated that this discrepancy is not due to different definitions of quiescent galaxies and velocity dispersion in observations and simulations. In Section \ref{sec:obs}, we discuss the implications of this discrepancy for deriving velocity dispersion functions.

The upper 95\% velocity dispersion limit at $\log (\Mstar / \Msun) = 9$ (our subhalo mass limit) corresponds to $\log \sigma_{*} \approx 1.7$. In other words, the velocity dispersion function we derive based on the subhalo sample with $\log (\Mstar / \Msun) = 9$ is complete to $\log \sigma_{*} = 1.7$. The velocity dispersion completeness limit is lower than the limit for observed velocity dispersion functions ($\sim 70-100~\kms$) based on the SDSS and MMT/Hectospec spectroscopic surveys for the local universe (e.g., \citealp{Choi07, Sohn17a, Sohn17b}).

\section{Velocity Dispersion Functions} \label{sec:result}

\begin{figure}
\centering
\includegraphics[scale=0.29]{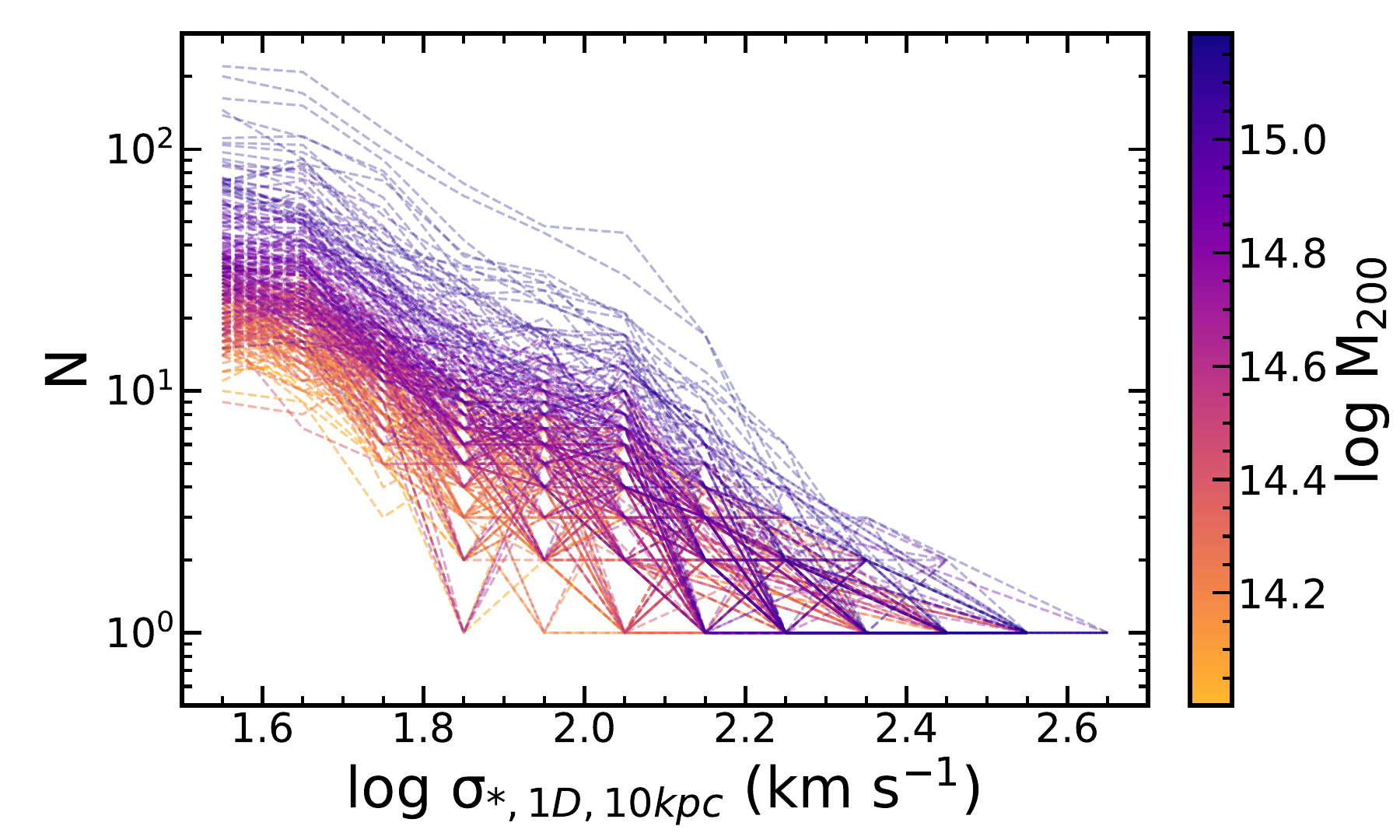}
\caption{Velocity dispersion function of subhalos in 280 massive clusters with $\log (M_{200} / \Msun) > 14$ in TNG300. Bluer colors display the velocity dispersion functions of more massive clusters. }
\label{fig:vdf}
\end{figure}

Derivation of the velocity dispersion function from simulations is straightforward. We simply count the number of member subhalos in each velocity dispersion bin. As Figure \ref{fig:msigma}(b) displays, the cluster subhalo samples with $\Mstar > 10^{9}~\Mstar$ are complete to $\log \sigma_{*} \sim 1.7$. Thus, measuring simulated velocity dispersion functions to this velocity dispersion limit requires no complex corrections. Figure \ref{fig:vdf} displays the velocity dispersion functions color-coded by cluster halo mass.

\begin{deluxetable*}{lcc}
\label{tab:bins}
\tablecaption{The Cluster Mass Distribution}
\tablecolumns{3}
\tablewidth{0pt}
\tablehead{\colhead{Mass Range} & \colhead{TNG300} & \colhead{HeCS-omnibus}}
\startdata
$15.0 < \log (M_{200} / \Msun) < 15.5$ &   3 & 13 \\
$14.8 < \log (M_{200} / \Msun) < 15.0$ &   6 & 35 \\
$14.6 < \log (M_{200} / \Msun) < 14.8$ &  13 & 48 \\
$14.4 < \log (M_{200} / \Msun) < 14.6$ &  40 & 54 \\
$14.2 < \log (M_{200} / \Msun) < 14.4$ &  78 & 27 \\
$14.0 < \log (M_{200} / \Msun) < 14.2$ & 140 & 22  
\enddata 
\end{deluxetable*}

\begin{figure*}
\centering
\includegraphics[scale=0.5]{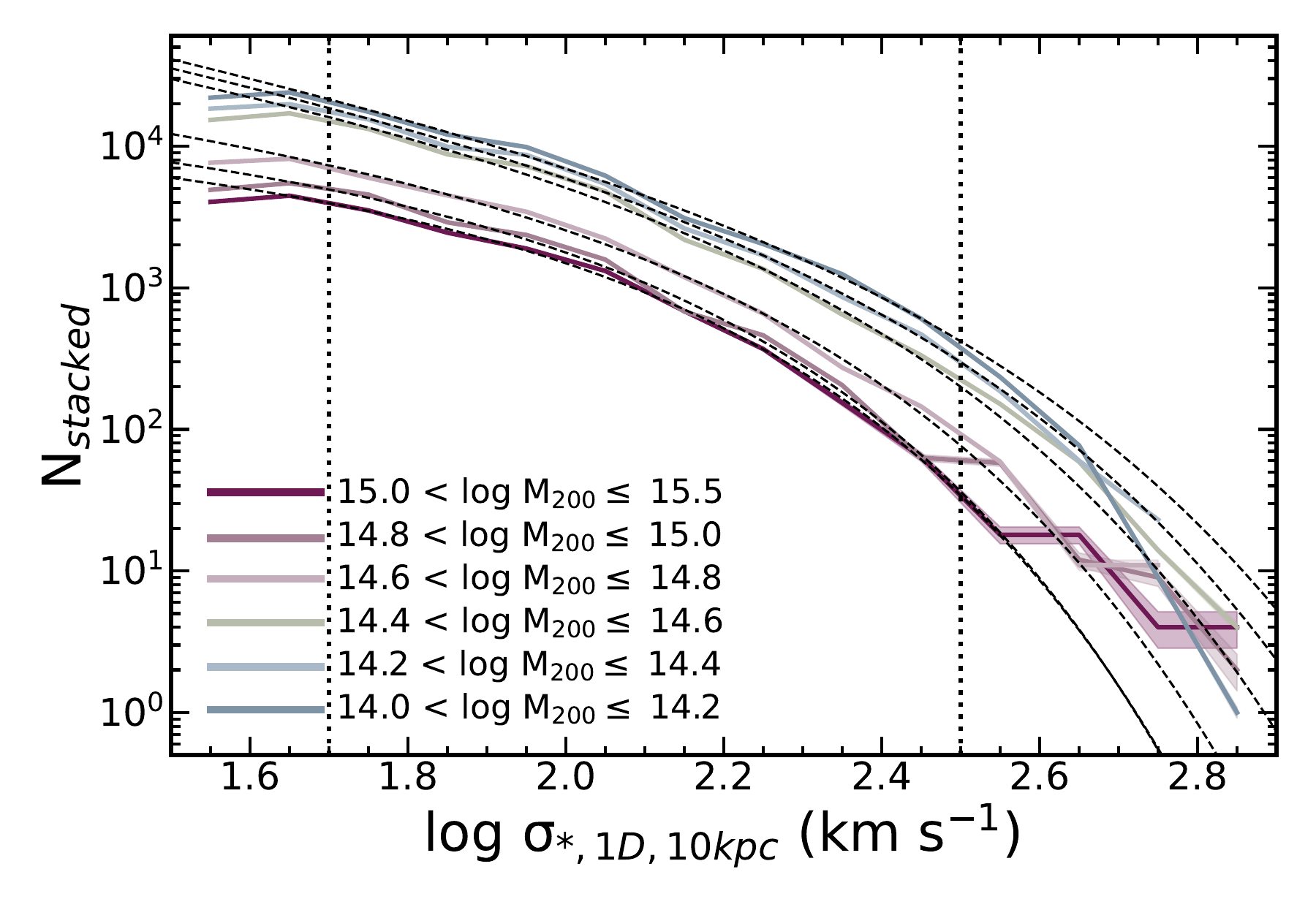}
\caption{Stacked velocity dispersion functions for clusters in six mass bins. Solid lines show the velocity dispersion function; the shaded regions show the uncertainties. For more massive systems, the number of clusters (darker color) per bin is smaller. Dashed lines display the best-fit Schechter functions. Vertical dotted lines mark the fitting range we used for Schechter function fits. } 
\label{fig:vdf_stack}
\end{figure*}

Figure \ref{fig:vdf_stack} shows the velocity dispersion functions based on stacked samples. We divide the cluster halos into six mass bins (see Table \ref{tab:bins}). We then count the number of subhalos within each bin and within $R_{200}$. Bluer color shows the velocity dispersion functions for the stacked sample of the more massive clusters. More massive stacked samples contain fewer subhalos because there are fewer very massive cluster halos. The shaded region indicates the Poisson noise in the velocity dispersion functions. 

We fit the stacked velocity dispersion functions with a Schechter function \citep{Schechter76} following \citet{Sohn20b}:
\begin{equation}
N_{\mathrm galaxy} = (\ln 10)\Phi^{*}10^{(\Sigma - \Sigma^{*})(1+\alpha)}\exp(-10^{(\Sigma-\Sigma^{*})}),
\end{equation}
where $\alpha$ is the slope, and $\Phi^{*}$ is the normalization, $\Sigma = \log \sigma_{*}$, and $\Sigma^{*} = \log \sigma^{*}_{*}$. Here, $\sigma^{*}$ is a characteristic stellar velocity dispersion where the velocity dispersion functions show a rapid change in slope. Previous studies often fit VDFs with a modified Schechter function (Eq. 2, \citealp{Choi07, Chae10, Montero17, Sohn17a, Sohn17b}). We do not use the modified Schechter function because the additional parameter that defines this function is poorly constrained. We compare the Schechter function fits with the observed VDFs in Section \ref{sec:obs}. Table \ref{tab:sch_fit} summarizes the best-fit Schechter function parameters for the simulated velocity dispersion functions. 

\begin{deluxetable*}{lccc}
\label{tab:sch_fit}
\tablecaption{Best-fit Schechter Function Parameters for Stacked Cluster Samples}
\tablecolumns{4}
\tablewidth{0pt}
\tablehead{\colhead{Mass Range} & \colhead{$\Phi^{*}$} & \colhead{$\alpha$} & \colhead{$\sigma^{*}$}}
\startdata
$15.0 < \log (M_{200} / \Msun) < 15.5$ & $3661 \pm  363$ & $-1.249 \pm 0.204$ & $61.97 \pm 1.09$ \\
$14.8 < \log (M_{200} / \Msun) < 15.0$ & $4587 \pm  882$ & $-1.303 \pm 0.402$ & $61.11 \pm 1.19$ \\
$14.6 < \log (M_{200} / \Msun) < 14.8$ & $4768 \pm 1411$ & $-1.587 \pm 0.327$ & $76.55 \pm 1.20$ \\
$14.4 < \log (M_{200} / \Msun) < 14.6$ & $5865 \pm 3172$ & $-1.951 \pm 0.353$ & $101.1 \pm 1.30$ \\
$14.2 < \log (M_{200} / \Msun) < 14.4$ & $4520 \pm 3202$ & $-2.081 \pm 0.353$ & $124.3 \pm 1.37$ \\
$14.0 < \log (M_{200} / \Msun) < 14.2$ & $3970 \pm 2797$ & $-2.135 \pm 0.300$ & $144.3 \pm 1.36$
\enddata 
\end{deluxetable*}

\section{Comparison with Other Velocity Dispersion Functions} \label{sec:comparison}

We determine the velocity dispersion functions based on the $\log (\Mstar / \Msun) > 9$ subhalos within TNG300 clusters. We first compare the cluster velocity dispersion functions with the general `field' velocity dispersion function also derived from TNG300 in Section \ref{sec:fld}. In Section \ref{sec:obs}, we compare simulated and observed velocity dispersion functions.

\subsection{Comparison with field velocity dispersion functions} \label{sec:fld}

We compare the cluster velocity dispersion function with the velocity dispersion function for the general field populations. \citet{Sohn24} compute the stellar velocity dispersion based on TNG300 subhalos with $\log (\Mstar / \Msun) > 9$. Among these subhalos, we select 104290 quiescent subhalos with sSFR $> 2 \times 10^{-11}$ yr$^{-1}$ over a volume of $\sim (300 {\mathrm Mpc})^{3}$. The selection is same as the selection for quiescent subhalos in clusters. We use these stellar velocity dispersions to construct the field velocity dispersion function. 

The field velocity dispersion function (i.e., the number of subhalos per unit volume) is significantly lower than the cluster velocity dispersion function. Thus, we scale the field velocity dispersion function to match the cluster velocity dispersion function at $\log \sigma_{*} = 2$ and then compare the shapes of the two functions. The normalization fraction corresponds to $\sim 223$. 

\begin{figure*}
\centering
\includegraphics[scale=0.5]{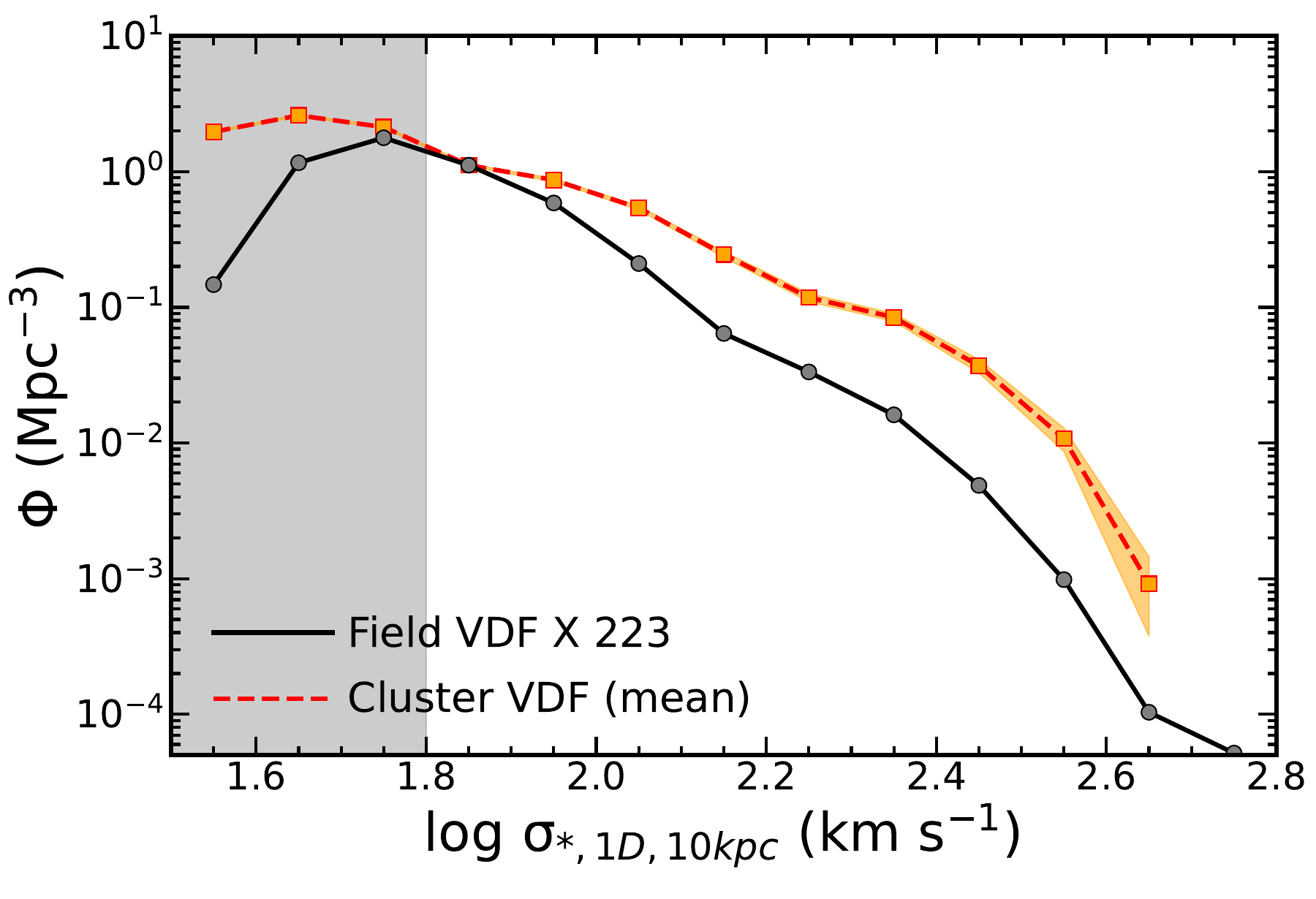}
\caption{Comparison between field (the black solid line) and cluster (the red-dashed line) velocity dispersion functions. The field velocity dispersion function represents all subhalos with $\log (\Mstar / \Msun) > 9$ and sSFR $> 2 \times 10^{-11}$ yr$^{-1}$. The cluster velocity dispersion function is the stacked velocity dispersion function for 280 clusters with $\log (M_{200} / \Msun) > 14$. To compare the shape of the two velocity dispersion functions, the field velocity dispersion function is scaled upward by $\sim 223$. The cluster velocity dispersion function exceeds the field for dispersions $\log \sigma_{*} > 2.2$. }
\label{fig:vdf_fld}
\end{figure*}

Figure \ref{fig:vdf_fld} compares the field (the black solid line) and cluster (the red dashed line) velocity dispersion functions. As in the observations \citep{Sohn17b}, the field velocity dispersion function lies below the cluster velocity dispersion function at large dispersion. 

The observed velocity dispersion functions show a similar contrast at large velocity dispersion. \citet{Sohn17b} compare the cluster velocity dispersion functions (for Coma and A2029) with the SDSS field velocity dispersion function. They demonstrate that the cluster velocity dispersion functions for Coma and A2029 significantly exceed the field when $\log \sigma > 2.2$. \citet{Sohn20a} also show that the velocity dispersion functions for nine clusters at $z > 0.2$ also exceed the field velocity dispersion function at large dispersion. 

\citet{Sohn17b} and \citet{Sohn20a} argue that the large dispersion excess of the cluster velocity dispersion function results from the preferential formation of massive galaxies in dense cluster environments. The comparison between field and cluster velocity dispersion functions in TNG300 supports this conjecture. 

\subsection{Comparison with observed cluster velocity dispersion functions}\label{sec:obs}

\citet{Sohn24} carefully explored simulated stellar velocity dispersions derived with different definitions to assess possible systematics. They demonstrate that the viewing axes, the velocity dispersion computation technique, and the resolution of simulations do not impact the velocity dispersion significantly. The systematic uncertainties resulting from different velocity dispersion definitions are smaller than the random errors. They also investigate the velocity dispersion dependence on the aperture of the cylindrical volume containing the member stellar particles. In general, the velocity dispersion measurements do not vary significantly with the aperture when $\log \sigma_{*} \sim 2.4$. For $\log \sigma_{*} > 2.4$, the stellar velocity dispersions are larger within a larger aperture because the stellar particles in the outer region of the galaxies tend to have larger velocity offsets. Thus the definition of the stellar velocity dispersions should not affect the comparison of observed and simulated velocity dispersion functions significantly. 

\begin{figure*}
\centering
\includegraphics[scale=0.5]{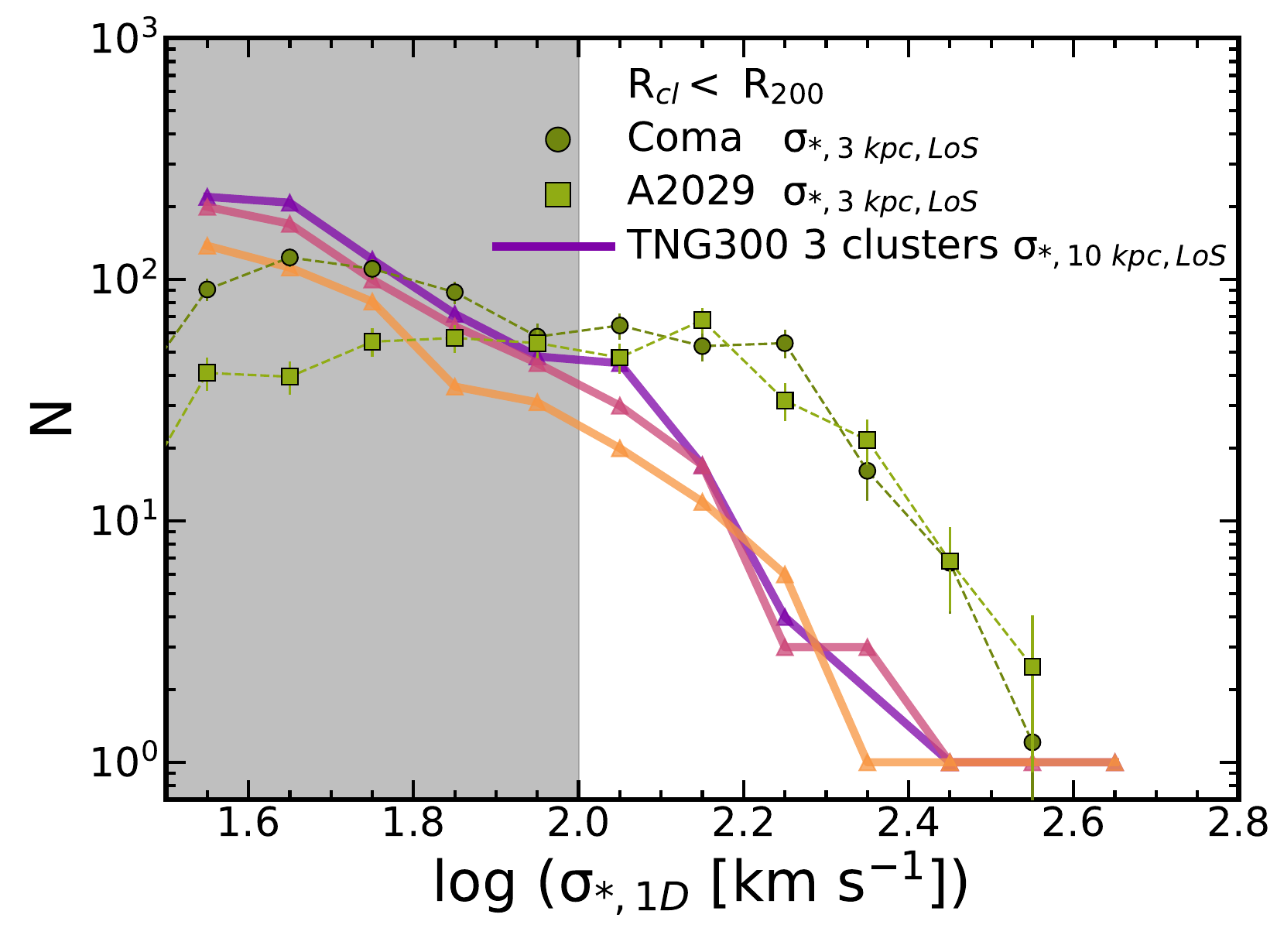}
\caption{Comparison between observed and simulated velocity dispersion functions. Green circles and squares show the observed velocity dispersion functions of Coma and A2029. Solid lines show the velocity dispersion function of the three most massive cluster halos ($\log (M_{200} / \Msun) > 15$) in TNG300. The shaded region indicates the velocity dispersion range where the incompleteness of the observed velocity dispersion function becomes significant.}
\label{fig:vdf_obs_aper}
\end{figure*}

Figure \ref{fig:vdf_obs_aper} compares the observed and simulated velocity dispersion functions. Green circles and squares display the Coma and A2029 velocity dispersion functions based on fiducial 3 kpc stellar velocity dispersion observations \citep{Sohn17a}. The gray shaded region shows the velocity dispersion range where the incompleteness of the observed velocity dispersion functions becomes significant. For comparison, we select the three most massive TNG clusters with $M_{200} > 10^{15}~\Msun$. Interestingly, the simulated VDFs are significantly below the observed VDFs for $\log \sigma_{*} > 2.0$. 

We derive the median of the three TNG cluster VDFs based on the stellar velocity dispersion measured within various apertures (3 kpc, 10 kpc, and the half-mass radius). As discussed in \citet{Sohn24}, stellar velocity dispersion measurements within different apertures do not affect the shape of the VDFs. Thus, the choice of the aperture is not a major cause of the difference between the observed and simulated VDFs in Figure \ref{fig:vdf_obs_aper}. 

The first issue in exploring the source of the difference between observed and simulated VDFs is the difference in the number of objects in simulated and observed clusters. Figure \ref{fig:nsub} shows that the number of simulated subhalos with $\log (\Mstar / \Msun) > 10$ in the TNG300 clusters is generally smaller than the observed number (HeCS-omnibus, \citealp{Sohn20a}). The TNG300 clusters generally contain 40\% fewer galaxies with stellar mass larger than $\log (\Mstar / \Msun) = 10$. This difference in counts contributes to the difference in the velocity dispersion functions for $\log \sigma_{*} > 2.0$. 

The second source of the discrepancy is the difference between the $\Mstar - \sigma_{*}$ relations for observed and simulated samples. The TNG300 subhalos display a tight relation between the stellar mass and the stellar velocity dispersions for $\log (\Mstar / \Msun) > 10$ (Figure \ref{fig:msigma}). However, the simulated $\Mstar - \sigma_{*}$ relation is systematically offset relative to the observed $\Mstar - \sigma_{*}$ relation. \citet{Sohn24} (their Figure 10) demonstrate that at a fixed stellar mass,  simulated velocity dispersions are 0.3 dex smaller than the stellar velocity dispersions derived from SDSS spectroscopy. This systematic offset appears at $\log (\Mstar / \Msun) = 10$, and the stellar velocity dispersions show a large scatter. Because the simulated velocity dispersions are systematically smaller than the observed velocity dispersions, the resulting velocity dispersion functions shift toward lower velocity dispersions. 

\begin{figure}
\centering
\includegraphics[scale=0.29]{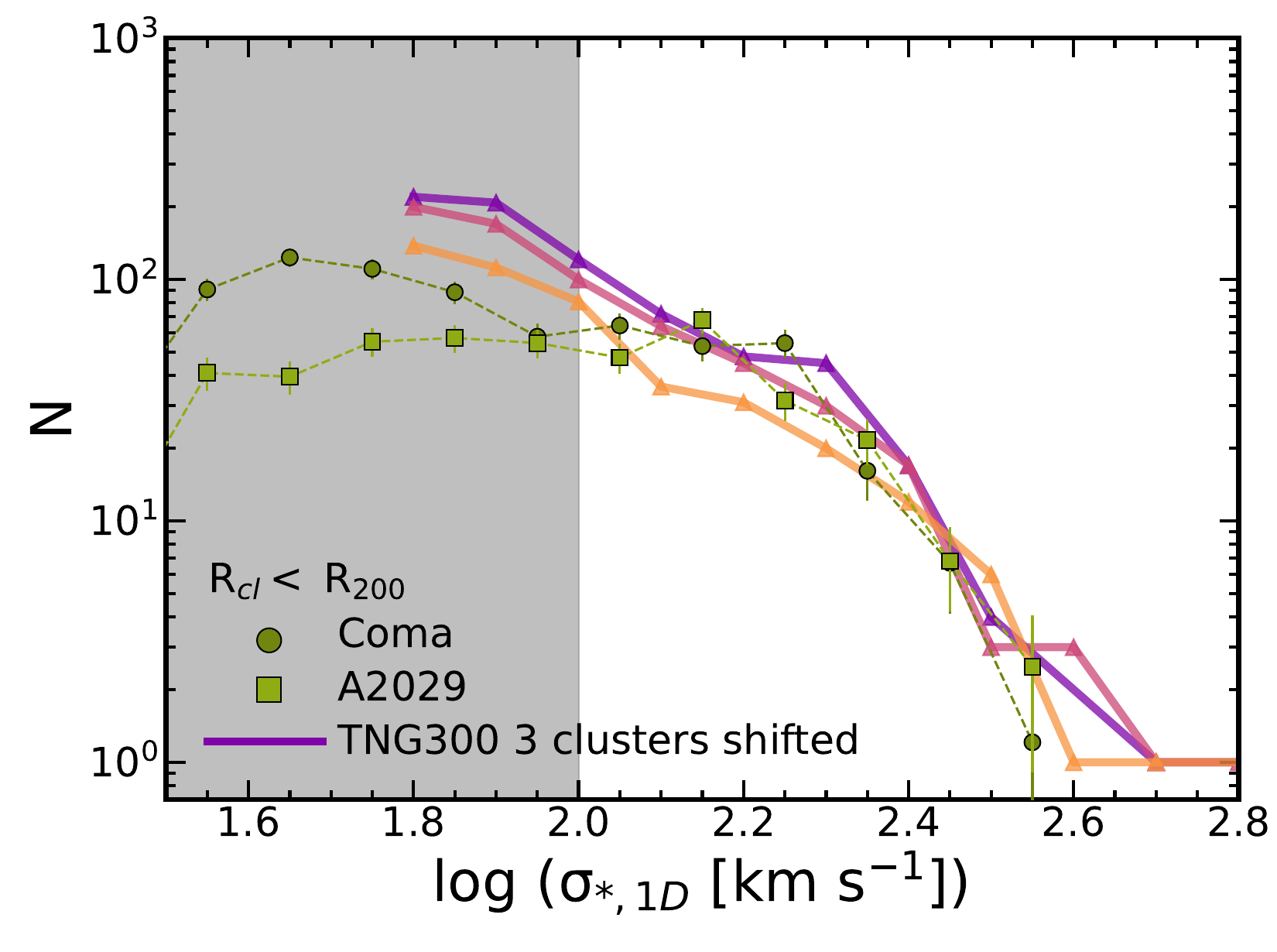}
\caption{Comparison between observed and simulated velocity dispersion functions. Green circles and squares show the observed velocity dispersion functions of Coma and A2029. Solid lines show the velocity dispersion function of the three most massive cluster halos ($\log (M_{200} / \Msun) > 15$) in TNG300, but shifted to match with the observed velocity dispersion functions. }
\label{fig:vdf_shift}
\end{figure}

We thus compare the slopes of the observed and simulated VDFs after shifting the simulated VDFs to correspond to the observations. In Figure \ref{fig:vdf_shift}, we shift the simulated velocity dispersions horizontally by 0.28 dex. The shift we apply is slightly larger than the offset in the $\Mstar - \sigma_{*}$ relation. This difference in shift accounts for the difference in the number of subhalos. 

The slopes of the shifted observed and simulated VDFs are consistent at $2.0 < \log \sigma_{*} < 2.6$. The consistent slopes of the VDFs suggest that TNG simulations predict the relative rates of massive galaxy formation despite the stellar velocity dispersion offsets.  

\section{CONCLUSION} \label{sec:conclusion}

We derive the stellar velocity dispersion functions for 280 massive cluster halos with $\log (M_{200} / \Msun) > 14$ in the IllustrisTNG 300 simulations. Following \citet{Sohn24}, we  select quiescent subhalos with specific star formation rates ($> 2 \times 10^{-11}$ yr$^{-1}$) and with $\log (\Mstar / \Msun) > 9$ within cluster halos. For the quiescent subhalos, we compute the stellar velocity dispersion in analogy with the stellar velocity dispersion derived from optical spectroscopy. We select stellar particles within a cylindrical volume that penetrates the center of the subhalos. Aperture sizes of the cylindrical volume ranging from 3 kpc to the half-mass radius do not affect the stellar velocity dispersions significantly. The subhalo sample with $\log (\Mstar / \Msun) > 9$ is complete to $\log \sigma_{*} \sim 1.7$. 

The cluster velocity dispersion functions for TNG300 clusters have a similar shape regardless of the cluster mass. The best-fit Schechter function parameters based on stacked velocity dispersion functions in six cluster mass bins are consistent with one another.  

We compare the cluster velocity dispersion function with the general field velocity dispersion function of quiescent subhalos. The stacked cluster velocity dispersion function is $\sim 200$ larger than the field velocity dispersion function reflecting the much larger density of clusters. The cluster velocity dispersion function at $\log \sigma_{*} > 2.17$  significantly exceeds the field velocity dispersion function. This difference is consistent with the difference between the observed field and cluster velocity dispersion functions \citep{Sohn17b}. 

We compare the TNG cluster velocity dispersion function with the observed velocity dispersion functions for two massive clusters, Coma and A2029 \citep{Sohn17a}. We select the three most massive simulated clusters with similarly massive ($\log (M_{200} / \Msun) > 15$) systems. The simulated cluster velocity dispersion functions lie substantially below compared to the observed functions. 

There are two main causes of the difference between the observed and simulated cluster velocity dispersion functions. First, the simulated clusters contain fewer ($\sim 60\%$) subhalos than the observed clusters to the same stellar mass limit. Second, the simulated stellar velocity dispersions are generally smaller than the observed stellar velocity dispersion at fixed stellar mass \citep{Sohn24}. These two effects combine to produce a substantially simulated velocity dispersion function that lies significantly below the observed velocity dispersion function. 

Although the simulated velocity dispersion functions differ from the observed velocity dispersion functions, the simulated velocity dispersion function still offers insights into the relative formation of massive subhaloes within clusters. After shifting the simulated velocity dispersion functions to match the observed velocity dispersion function, the slopes at $\log \sigma_{*} > 2$ are essentially identical. Thus, the TNG300 simulations predict the relative distribution of velocity dispersions that is consistent with the observations. The simulations support the idea that massive subhalo formation is more efficient in the dense cluster environment.

The velocity dispersion function we derive from TNG300 is a potential probe of the mass or velocity dispersion distribution of underlying dark matter subhalos. \citet{Sohn24} demonstrate that the stellar velocity dispersions have a relation with the dark matter velocity dispersions. Using Illustris-1 simulation, \citet{Zahid17} show that the stellar velocity dispersion of the subhalos (more specifically satellite subhalos) is connected to the dark matter velocity dispersion (or mass) at the epoch of accretion onto the cluster halo. Further investigation on the relation between stellar and dark matter velocity dispersions based on the TNG300 or other simulations will establish the foundation for using future observed velocity dispersion functions to tracing the underlying dark matter velocity dispersion (mass) distributions as a function of cosmic time.

\begin{acknowledgments}
We thank Scott Kanyon, Ken Rines, Antonaldo Diaferio, Ivana Damjanov, and Michele Pizzardo for their helpful discussions and comments while preparing this manuscript. JS is supported by the National Research Foundation of Korea (NRF) grant funded by the Korean government (MSIT) (RS-2023-00210597). This work was supported by the New Faculty Startup Fund from Seoul National University. This work was also supported by the Global-LAMP Program of the National Research Foundation of Korea (NRF) grant funded by the Ministry of Education (No. RS-2023-00301976). JB acknowledges support from NSF grant AST-2153201. 
\end{acknowledgments}

\software{astropy \citep{astropy13, astropy18}, scipy \citep{SciPy}, MatPlotlib \citep{Matplotlib}, h5py}

\bibliography{ms}{}

\begin{thebibliography}{}
\expandafter\ifx\csname natexlab\endcsname\relax\def\natexlab#1{#1}\fi
\providecommand{\url}[1]{\href{#1}{#1}}
\providecommand{\dodoi}[1]{doi:~\href{http://doi.org/#1}{\nolinkurl{#1}}}
\providecommand{\doeprint}[1]{\href{http://ascl.net/#1}{\nolinkurl{http://ascl.net/#1}}}
\providecommand{\doarXiv}[1]{\href{https://arxiv.org/abs/#1}{\nolinkurl{https://arxiv.org/abs/#1}}}

\bibitem[{{Astropy Collaboration} {et~al.}(2013){Astropy Collaboration},
  {Robitaille}, {Tollerud}, {Greenfield}, {Droettboom}, {Bray}, {Aldcroft},
  {Davis}, {Ginsburg}, {Price-Whelan}, {Kerzendorf}, {Conley}, {Crighton},
  {Barbary}, {Muna}, {Ferguson}, {Grollier}, {Parikh}, {Nair}, {Unther},
  {Deil}, {Woillez}, {Conseil}, {Kramer}, {Turner}, {Singer}, {Fox}, {Weaver},
  {Zabalza}, {Edwards}, {Azalee Bostroem}, {Burke}, {Casey}, {Crawford},
  {Dencheva}, {Ely}, {Jenness}, {Labrie}, {Lim}, {Pierfederici}, {Pontzen},
  {Ptak}, {Refsdal}, {Servillat}, \& {Streicher}}]{astropy13}
{Astropy Collaboration}, {Robitaille}, T.~P., {Tollerud}, E.~J., {et~al.} 2013,
  \aap, 558, A33, \dodoi{10.1051/0004-6361/201322068}

\bibitem[{{Astropy Collaboration} {et~al.}(2018){Astropy Collaboration},
  {Price-Whelan}, {Sip{\H{o}}cz}, {G{\"u}nther}, {Lim}, {Crawford}, {Conseil},
  {Shupe}, {Craig}, {Dencheva}, {Ginsburg}, {VanderPlas}, {Bradley},
  {P{\'e}rez-Su{\'a}rez}, {de Val-Borro}, {Aldcroft}, {Cruz}, {Robitaille},
  {Tollerud}, {Ardelean}, {Babej}, {Bach}, {Bachetti}, {Bakanov}, {Bamford},
  {Barentsen}, {Barmby}, {Baumbach}, {Berry}, {Biscani}, {Boquien}, {Bostroem},
  {Bouma}, {Brammer}, {Bray}, {Breytenbach}, {Buddelmeijer}, {Burke},
  {Calderone}, {Cano Rodr{\'\i}guez}, {Cara}, {Cardoso}, {Cheedella}, {Copin},
  {Corrales}, {Crichton}, {D'Avella}, {Deil}, {Depagne}, {Dietrich}, {Donath},
  {Droettboom}, {Earl}, {Erben}, {Fabbro}, {Ferreira}, {Finethy}, {Fox},
  {Garrison}, {Gibbons}, {Goldstein}, {Gommers}, {Greco}, {Greenfield},
  {Groener}, {Grollier}, {Hagen}, {Hirst}, {Homeier}, {Horton}, {Hosseinzadeh},
  {Hu}, {Hunkeler}, {Ivezi{\'c}}, {Jain}, {Jenness}, {Kanarek}, {Kendrew},
  {Kern}, {Kerzendorf}, {Khvalko}, {King}, {Kirkby}, {Kulkarni}, {Kumar},
  {Lee}, {Lenz}, {Littlefair}, {Ma}, {Macleod}, {Mastropietro}, {McCully},
  {Montagnac}, {Morris}, {Mueller}, {Mumford}, {Muna}, {Murphy}, {Nelson},
  {Nguyen}, {Ninan}, {N{\"o}the}, {Ogaz}, {Oh}, {Parejko}, {Parley}, {Pascual},
  {Patil}, {Patil}, {Plunkett}, {Prochaska}, {Rastogi}, {Reddy Janga},
  {Sabater}, {Sakurikar}, {Seifert}, {Sherbert}, {Sherwood-Taylor}, {Shih},
  {Sick}, {Silbiger}, {Singanamalla}, {Singer}, {Sladen}, {Sooley},
  {Sornarajah}, {Streicher}, {Teuben}, {Thomas}, {Tremblay}, {Turner},
  {Terr{\'o}n}, {van Kerkwijk}, {de la Vega}, {Watkins}, {Weaver}, {Whitmore},
  {Woillez}, {Zabalza}, \& {Astropy Contributors}}]{astropy18}
{Astropy Collaboration}, {Price-Whelan}, A.~M., {Sip{\H{o}}cz}, B.~M., {et~al.}
  2018, \aj, 156, 123, \dodoi{10.3847/1538-3881/aabc4f}

\bibitem[{{Belli} {et~al.}(2014){Belli}, {Newman}, \& {Ellis}}]{Belli14}
{Belli}, S., {Newman}, A.~B., \& {Ellis}, R.~S. 2014, \apj, 783, 117,
  \dodoi{10.1088/0004-637X/783/2/117}

\bibitem[{{Bernardi} {et~al.}(2017){Bernardi}, {Meert}, {Sheth}, {Fischer},
  {Huertas-Company}, {Maraston}, {Shankar}, \& {Vikram}}]{Bernardi17}
{Bernardi}, M., {Meert}, A., {Sheth}, R.~K., {et~al.} 2017, \mnras, 467, 2217,
  \dodoi{10.1093/mnras/stx176}

\bibitem[{{Bernardi} {et~al.}(2013){Bernardi}, {Meert}, {Sheth}, {Vikram},
  {Huertas-Company}, {Mei}, \& {Shankar}}]{Bernardi13}
---. 2013, \mnras, 436, 697, \dodoi{10.1093/mnras/stt1607}

\bibitem[{{Bernardi} {et~al.}(2010){Bernardi}, {Shankar}, {Hyde}, {Mei},
  {Marulli}, \& {Sheth}}]{Bernardi10}
{Bernardi}, M., {Shankar}, F., {Hyde}, J.~B., {et~al.} 2010, \mnras, 404, 2087,
  \dodoi{10.1111/j.1365-2966.2010.16425.x}

\bibitem[{{Bogd{\'a}n} \& {Goulding}(2015)}]{Bogdan15}
{Bogd{\'a}n}, {\'A}., \& {Goulding}, A.~D. 2015, \apj, 800, 124,
  \dodoi{10.1088/0004-637X/800/2/124}

\bibitem[{{Cannarozzo} {et~al.}(2020){Cannarozzo}, {Sonnenfeld}, \&
  {Nipoti}}]{Cannarozzo20}
{Cannarozzo}, C., {Sonnenfeld}, A., \& {Nipoti}, C. 2020, \mnras, 498, 1101,
  \dodoi{10.1093/mnras/staa2147}

\bibitem[{{Cappellari}(2016)}]{Cappellari16}
{Cappellari}, M. 2016, \araa, 54, 597,
  \dodoi{10.1146/annurev-astro-082214-122432}

\bibitem[{{Cappellari} {et~al.}(2013){Cappellari}, {McDermid}, {Alatalo},
  {Blitz}, {Bois}, {Bournaud}, {Bureau}, {Crocker}, {Davies}, {Davis}, {de
  Zeeuw}, {Duc}, {Emsellem}, {Khochfar}, {Krajnovi{\'c}}, {Kuntschner},
  {Morganti}, {Naab}, {Oosterloo}, {Sarzi}, {Scott}, {Serra}, {Weijmans}, \&
  {Young}}]{Cappellari13}
{Cappellari}, M., {McDermid}, R.~M., {Alatalo}, K., {et~al.} 2013, \mnras, 432,
  1862, \dodoi{10.1093/mnras/stt644}

\bibitem[{{Chae}(2010)}]{Chae10}
{Chae}, K.-H. 2010, \mnras, 402, 2031, \dodoi{10.1111/j.1365-2966.2009.16073.x}

\bibitem[{{Choi} {et~al.}(2007){Choi}, {Park}, \& {Vogeley}}]{Choi07}
{Choi}, Y.-Y., {Park}, C., \& {Vogeley}, M.~S. 2007, \apj, 658, 884,
  \dodoi{10.1086/511060}

\bibitem[{{Damjanov} {et~al.}(2022){Damjanov}, {Sohn}, {Utsumi}, {Geller}, \&
  {Dell'Antonio}}]{Damjanov22}
{Damjanov}, I., {Sohn}, J., {Utsumi}, Y., {Geller}, M.~J., \& {Dell'Antonio},
  I. 2022, \apj, 929, 61, \dodoi{10.3847/1538-4357/ac54bd}

\bibitem[{{Damjanov} {et~al.}(2018){Damjanov}, {Zahid}, {Geller}, {Fabricant},
  \& {Hwang}}]{Damjanov18}
{Damjanov}, I., {Zahid}, H.~J., {Geller}, M.~J., {Fabricant}, D.~G., \&
  {Hwang}, H.~S. 2018, \apjs, 234, 21, \dodoi{10.3847/1538-4365/aaa01c}

\bibitem[{{Damjanov} {et~al.}(2019){Damjanov}, {Zahid}, {Geller}, {Utsumi},
  {Sohn}, \& {Souchereau}}]{Damjanov19}
{Damjanov}, I., {Zahid}, H.~J., {Geller}, M.~J., {et~al.} 2019, \apj, 872, 91,
  \dodoi{10.3847/1538-4357/aaf97d}

\bibitem[{{Djorgovski} \& {Davis}(1987)}]{Djorgovski87}
{Djorgovski}, S., \& {Davis}, M. 1987, \apj, 313, 59, \dodoi{10.1086/164948}

\bibitem[{{Faber} \& {Jackson}(1976)}]{Faber76}
{Faber}, S.~M., \& {Jackson}, R.~E. 1976, \apj, 204, 668,
  \dodoi{10.1086/154215}

\bibitem[{{Huchra} \& {Geller}(1982)}]{Huchra82}
{Huchra}, J.~P., \& {Geller}, M.~J. 1982, \apj, 257, 423,
  \dodoi{10.1086/160000}

\bibitem[{Hunter(2007)}]{Matplotlib}
Hunter, J.~D. 2007, Computing in Science \& Engineering, 9, 90,
  \dodoi{10.1109/MCSE.2007.55}

\bibitem[{{Hyde} \& {Bernardi}(2009)}]{Hyde09}
{Hyde}, J.~B., \& {Bernardi}, M. 2009, \mnras, 396, 1171,
  \dodoi{10.1111/j.1365-2966.2009.14783.x}

\bibitem[{{Li} {et~al.}(2022){Li}, {Huang}, {Leauthaud}, {Moustakas},
  {Danieli}, {Greene}, {Abraham}, {Ardila}, {Kado-Fong}, {Lokhorst}, {Lupton},
  \& {Price}}]{Li22}
{Li}, J., {Huang}, S., {Leauthaud}, A., {et~al.} 2022, \mnras, 515, 5335,
  \dodoi{10.1093/mnras/stac2121}

\bibitem[{{Marinacci} {et~al.}(2018){Marinacci}, {Vogelsberger}, {Pakmor},
  {Torrey}, {Springel}, {Hernquist}, {Nelson}, {Weinberger}, {Pillepich},
  {Naiman}, \& {Genel}}]{Marinacci18}
{Marinacci}, F., {Vogelsberger}, M., {Pakmor}, R., {et~al.} 2018, \mnras, 480,
  5113, \dodoi{10.1093/mnras/sty2206}

\bibitem[{{Miller} {et~al.}(2021){Miller}, {van Dokkum}, {Danieli}, {Li},
  {Abraham}, {Conroy}, {Gilhuly}, {Greco}, {Liu}, {Lokhorst}, \&
  {Merritt}}]{Miller21}
{Miller}, T.~B., {van Dokkum}, P., {Danieli}, S., {et~al.} 2021, \apj, 909, 74,
  \dodoi{10.3847/1538-4357/abd7f8}

\bibitem[{{Montero-Dorta} {et~al.}(2017){Montero-Dorta}, {Bolton}, \&
  {Shu}}]{Montero17}
{Montero-Dorta}, A.~D., {Bolton}, A.~S., \& {Shu}, Y. 2017, \mnras, 468, 47,
  \dodoi{10.1093/mnras/stx321}

\bibitem[{{Moresco} {et~al.}(2013){Moresco}, {Pozzetti}, {Cimatti}, {Zamorani},
  {Bolzonella}, {Lamareille}, {Mignoli}, {Zucca}, {Lilly}, {Carollo},
  {Contini}, {Kneib}, {Le F{\`e}vre}, {Mainieri}, {Renzini}, {Scodeggio},
  {Bardelli}, {Bongiorno}, {Caputi}, {Cucciati}, {de la Torre}, {de Ravel},
  {Franzetti}, {Garilli}, {Iovino}, {Kampczyk}, {Knobel}, {Kova{\v{c}}}, {Le
  Borgne}, {Le Brun}, {Maier}, {Pell{\'o}}, {Peng}, {Perez-Montero},
  {Presotto}, {Silverman}, {Tanaka}, {Tasca}, {Tresse}, {Vergani}, {Barnes},
  {Bordoloi}, {Cappi}, {Diener}, {Koekemoer}, {Le Floc'h}, {L{\'o}pez-Sanjuan},
  {McCracken}, {Nair}, {Oesch}, {Scarlata}, {Scoville}, \&
  {Welikala}}]{Moresco13}
{Moresco}, M., {Pozzetti}, L., {Cimatti}, A., {et~al.} 2013, \aap, 558, A61,
  \dodoi{10.1051/0004-6361/201321797}

\bibitem[{{Munari} {et~al.}(2016){Munari}, {Grillo}, {De Lucia}, {Biviano},
  {Annunziatella}, {Borgani}, {Lombardi}, {Mercurio}, \& {Rosati}}]{Munari16}
{Munari}, E., {Grillo}, C., {De Lucia}, G., {et~al.} 2016, \apjl, 827, L5,
  \dodoi{10.3847/2041-8205/827/1/L5}

\bibitem[{{Naiman} {et~al.}(2018){Naiman}, {Pillepich}, {Springel},
  {Ramirez-Ruiz}, {Torrey}, {Vogelsberger}, {Pakmor}, {Nelson}, {Marinacci},
  {Hernquist}, {Weinberger}, \& {Genel}}]{Naiman18}
{Naiman}, J.~P., {Pillepich}, A., {Springel}, V., {et~al.} 2018, \mnras, 477,
  1206, \dodoi{10.1093/mnras/sty618}

\bibitem[{{Nelson} {et~al.}(2015){Nelson}, {Pillepich}, {Genel},
  {Vogelsberger}, {Springel}, {Torrey}, {Rodriguez-Gomez}, {Sijacki}, {Snyder},
  {Griffen}, {Marinacci}, {Blecha}, {Sales}, {Xu}, \& {Hernquist}}]{Nelson15}
{Nelson}, D., {Pillepich}, A., {Genel}, S., {et~al.} 2015, Astronomy and
  Computing, 13, 12, \dodoi{10.1016/j.ascom.2015.09.003}

\bibitem[{{Nelson} {et~al.}(2018){Nelson}, {Pillepich}, {Springel},
  {Weinberger}, {Hernquist}, {Pakmor}, {Genel}, {Torrey}, {Vogelsberger},
  {Kauffmann}, {Marinacci}, \& {Naiman}}]{Nelson18}
{Nelson}, D., {Pillepich}, A., {Springel}, V., {et~al.} 2018, \mnras, 475, 624,
  \dodoi{10.1093/mnras/stx3040}

\bibitem[{{Nelson} {et~al.}(2019){Nelson}, {Springel}, {Pillepich},
  {Rodriguez-Gomez}, {Torrey}, {Genel}, {Vogelsberger}, {Pakmor}, {Marinacci},
  {Weinberger}, {Kelley}, {Lovell}, {Diemer}, \& {Hernquist}}]{Nelson19}
{Nelson}, D., {Springel}, V., {Pillepich}, A., {et~al.} 2019, Computational
  Astrophysics and Cosmology, 6, 2, \dodoi{10.1186/s40668-019-0028-x}

\bibitem[{{Pillepich} {et~al.}(2018){Pillepich}, {Springel}, {Nelson}, {Genel},
  {Naiman}, {Pakmor}, {Hernquist}, {Torrey}, {Vogelsberger}, {Weinberger}, \&
  {Marinacci}}]{Pillepich18}
{Pillepich}, A., {Springel}, V., {Nelson}, D., {et~al.} 2018, \mnras, 473,
  4077, \dodoi{10.1093/mnras/stx2656}

\bibitem[{{Planck Collaboration} {et~al.}(2016){Planck Collaboration}, {Ade},
  {Aghanim}, {Arnaud}, {Ashdown}, {Aumont}, {Baccigalupi}, {Banday},
  {Barreiro}, {Bartlett}, {Bartolo}, {Battaner}, {Battye}, {Benabed},
  {Beno{\^\i}t}, {Benoit-L{\'e}vy}, {Bernard}, {Bersanelli}, {Bielewicz},
  {Bock}, {Bonaldi}, {Bonavera}, {Bond}, {Borrill}, {Bouchet}, {Boulanger},
  {Bucher}, {Burigana}, {Butler}, {Calabrese}, {Cardoso}, {Catalano},
  {Challinor}, {Chamballu}, {Chary}, {Chiang}, {Chluba}, {Christensen},
  {Church}, {Clements}, {Colombi}, {Colombo}, {Combet}, {Coulais}, {Crill},
  {Curto}, {Cuttaia}, {Danese}, {Davies}, {Davis}, {de Bernardis}, {de Rosa},
  {de Zotti}, {Delabrouille}, {D{\'e}sert}, {Di Valentino}, {Dickinson},
  {Diego}, {Dolag}, {Dole}, {Donzelli}, {Dor{\'e}}, {Douspis}, {Ducout},
  {Dunkley}, {Dupac}, {Efstathiou}, {Elsner}, {En{\ss}lin}, {Eriksen},
  {Farhang}, {Fergusson}, {Finelli}, {Forni}, {Frailis}, {Fraisse},
  {Franceschi}, {Frejsel}, {Galeotta}, {Galli}, {Ganga}, {Gauthier}, {Gerbino},
  {Ghosh}, {Giard}, {Giraud-H{\'e}raud}, {Giusarma}, {Gjerl{\o}w},
  {Gonz{\'a}lez-Nuevo}, {G{\'o}rski}, {Gratton}, {Gregorio}, {Gruppuso},
  {Gudmundsson}, {Hamann}, {Hansen}, {Hanson}, {Harrison}, {Helou},
  {Henrot-Versill{\'e}}, {Hern{\'a}ndez-Monteagudo}, {Herranz}, {Hildebrandt},
  {Hivon}, {Hobson}, {Holmes}, {Hornstrup}, {Hovest}, {Huang}, {Huffenberger},
  {Hurier}, {Jaffe}, {Jaffe}, {Jones}, {Juvela}, {Keih{\"a}nen}, {Keskitalo},
  {Kisner}, {Kneissl}, {Knoche}, {Knox}, {Kunz}, {Kurki-Suonio}, {Lagache},
  {L{\"a}hteenm{\"a}ki}, {Lamarre}, {Lasenby}, {Lattanzi}, {Lawrence}, {Leahy},
  {Leonardi}, {Lesgourgues}, {Levrier}, {Lewis}, {Liguori}, {Lilje},
  {Linden-V{\o}rnle}, {L{\'o}pez-Caniego}, {Lubin}, {Mac{\'\i}as-P{\'e}rez},
  {Maggio}, {Maino}, {Mandolesi}, {Mangilli}, {Marchini}, {Maris}, {Martin},
  {Martinelli}, {Mart{\'\i}nez-Gonz{\'a}lez}, {Masi}, {Matarrese}, {McGehee},
  {Meinhold}, {Melchiorri}, {Melin}, {Mendes}, {Mennella}, {Migliaccio},
  {Millea}, {Mitra}, {Miville-Desch{\^e}nes}, {Moneti}, {Montier}, {Morgante},
  {Mortlock}, {Moss}, {Munshi}, {Murphy}, {Naselsky}, {Nati}, {Natoli},
  {Netterfield}, {N{\o}rgaard-Nielsen}, {Noviello}, {Novikov}, {Novikov},
  {Oxborrow}, {Paci}, {Pagano}, {Pajot}, {Paladini}, {Paoletti}, {Partridge},
  {Pasian}, {Patanchon}, {Pearson}, {Perdereau}, {Perotto}, {Perrotta},
  {Pettorino}, {Piacentini}, {Piat}, {Pierpaoli}, {Pietrobon}, {Plaszczynski},
  {Pointecouteau}, {Polenta}, {Popa}, {Pratt}, {Pr{\'e}zeau}, {Prunet},
  {Puget}, {Rachen}, {Reach}, {Rebolo}, {Reinecke}, {Remazeilles}, {Renault},
  {Renzi}, {Ristorcelli}, {Rocha}, {Rosset}, {Rossetti}, {Roudier},
  {Rouill{\'e} d'Orfeuil}, {Rowan-Robinson}, {Rubi{\~n}o-Mart{\'\i}n},
  {Rusholme}, {Said}, {Salvatelli}, {Salvati}, {Sandri}, {Santos},
  {Savelainen}, {Savini}, {Scott}, {Seiffert}, {Serra}, {Shellard}, {Spencer},
  {Spinelli}, {Stolyarov}, {Stompor}, {Sudiwala}, {Sunyaev}, {Sutton},
  {Suur-Uski}, {Sygnet}, {Tauber}, {Terenzi}, {Toffolatti}, {Tomasi},
  {Tristram}, {Trombetti}, {Tucci}, {Tuovinen}, {T{\"u}rler}, {Umana},
  {Valenziano}, {Valiviita}, {Van Tent}, {Vielva}, {Villa}, {Wade}, {Wandelt},
  {Wehus}, {White}, {White}, {Wilkinson}, {Yvon}, {Zacchei}, \&
  {Zonca}}]{Planck16}
{Planck Collaboration}, {Ade}, P.~A.~R., {Aghanim}, N., {et~al.} 2016, \aap,
  594, A13, \dodoi{10.1051/0004-6361/201525830}

\bibitem[{{Schechter}(1976)}]{Schechter76}
{Schechter}, P. 1976, \apj, 203, 297, \dodoi{10.1086/154079}

\bibitem[{{Schechter}(2015)}]{Schechter15}
{Schechter}, P.~L. 2015, arXiv e-prints, arXiv:1508.02358,
  \dodoi{10.48550/arXiv.1508.02358}

\bibitem[{{Seo} {et~al.}(2020){Seo}, {Sohn}, \& {Lee}}]{Seo20}
{Seo}, G., {Sohn}, J., \& {Lee}, M.~G. 2020, \apj, 903, 130,
  \dodoi{10.3847/1538-4357/abbd92}

\bibitem[{{Sheth} {et~al.}(2003){Sheth}, {Bernardi}, {Schechter}, {Burles},
  {Eisenstein}, {Finkbeiner}, {Frieman}, {Lupton}, {Schlegel}, {Subbarao},
  {Shimasaku}, {Bahcall}, {Brinkmann}, \& {Ivezi{\'c}}}]{Sheth03}
{Sheth}, R.~K., {Bernardi}, M., {Schechter}, P.~L., {et~al.} 2003, \apj, 594,
  225, \dodoi{10.1086/376794}

\bibitem[{{Sohn} {et~al.}(2020{\natexlab{a}}){Sohn}, {Fabricant}, {Geller},
  {Hwang}, \& {Diaferio}}]{Sohn20a}
{Sohn}, J., {Fabricant}, D.~G., {Geller}, M.~J., {Hwang}, H.~S., \& {Diaferio},
  A. 2020{\natexlab{a}}, \apj, 902, 17, \dodoi{10.3847/1538-4357/abb23b}

\bibitem[{{Sohn} {et~al.}(2024){Sohn}, {Geller}, {Borrow}, \&
  {Vogelsberger}}]{Sohn24}
{Sohn}, J., {Geller}, M.~J., {Borrow}, J., \& {Vogelsberger}, M. 2024, \apj,
  964, 178, \dodoi{10.3847/1538-4357/ad2c0a}

\bibitem[{{Sohn} {et~al.}(2020{\natexlab{b}}){Sohn}, {Geller}, {Diaferio}, \&
  {Rines}}]{Sohn20b}
{Sohn}, J., {Geller}, M.~J., {Diaferio}, A., \& {Rines}, K.~J.
  2020{\natexlab{b}}, \apj, 891, 129, \dodoi{10.3847/1538-4357/ab6e6a}

\bibitem[{{Sohn} {et~al.}(2021){Sohn}, {Geller}, {Hwang}, {Diaferio}, {Rines},
  \& {Utsumi}}]{Sohn21}
{Sohn}, J., {Geller}, M.~J., {Hwang}, H.~S., {et~al.} 2021, \apj, 923, 143,
  \dodoi{10.3847/1538-4357/ac29c3}

\bibitem[{{Sohn} {et~al.}(2017{\natexlab{a}}){Sohn}, {Geller}, {Zahid},
  {Fabricant}, {Diaferio}, \& {Rines}}]{Sohn17a}
{Sohn}, J., {Geller}, M.~J., {Zahid}, H.~J., {et~al.} 2017{\natexlab{a}},
  \apjs, 229, 20, \dodoi{10.3847/1538-4365/aa653e}

\bibitem[{{Sohn} {et~al.}(2017{\natexlab{b}}){Sohn}, {Zahid}, \&
  {Geller}}]{Sohn17b}
{Sohn}, J., {Zahid}, H.~J., \& {Geller}, M.~J. 2017{\natexlab{b}}, \apj, 845,
  73, \dodoi{10.3847/1538-4357/aa7de3}

\bibitem[{{Springel} {et~al.}(2001){Springel}, {White}, {Tormen}, \&
  {Kauffmann}}]{Springel01}
{Springel}, V., {White}, S. D.~M., {Tormen}, G., \& {Kauffmann}, G. 2001,
  \mnras, 328, 726, \dodoi{10.1046/j.1365-8711.2001.04912.x}

\bibitem[{{Springel} {et~al.}(2018){Springel}, {Pakmor}, {Pillepich},
  {Weinberger}, {Nelson}, {Hernquist}, {Vogelsberger}, {Genel}, {Torrey},
  {Marinacci}, \& {Naiman}}]{Springel18}
{Springel}, V., {Pakmor}, R., {Pillepich}, A., {et~al.} 2018, \mnras, 475, 676,
  \dodoi{10.1093/mnras/stx3304}

\bibitem[{{Utsumi} {et~al.}(2020){Utsumi}, {Geller}, {Zahid}, {Sohn},
  {Dell'Antonio}, {Kawanomoto}, {Komiyama}, {Koshida}, \&
  {Miyazaki}}]{Utsumi20}
{Utsumi}, Y., {Geller}, M.~J., {Zahid}, H.~J., {et~al.} 2020, \apj, 900, 50,
  \dodoi{10.3847/1538-4357/aba61c}

\bibitem[{{van Uitert} {et~al.}(2013){van Uitert}, {Hoekstra}, {Franx},
  {Gilbank}, {Gladders}, \& {Yee}}]{vanuitert13}
{van Uitert}, E., {Hoekstra}, H., {Franx}, M., {et~al.} 2013, \aap, 549, A7,
  \dodoi{10.1051/0004-6361/201220439}

\bibitem[{Virtanen {et~al.}(2020)Virtanen, Gommers, Oliphant, Haberland, Reddy,
  Cournapeau, Burovski, Peterson, Weckesser, Bright, {van der Walt}, Brett,
  Wilson, Millman, Mayorov, Nelson, Jones, Kern, Larson, Carey, Polat, Feng,
  Moore, {VanderPlas}, Laxalde, Perktold, Cimrman, Henriksen, Quintero, Harris,
  Archibald, Ribeiro, Pedregosa, {van Mulbregt}, \& {SciPy 1.0
  Contributors}}]{SciPy}
Virtanen, P., Gommers, R., Oliphant, T.~E., {et~al.} 2020, Nature Methods, 17,
  261, \dodoi{10.1038/s41592-019-0686-2}

\bibitem[{{Vogelsberger} {et~al.}(2014){Vogelsberger}, {Genel}, {Springel},
  {Torrey}, {Sijacki}, {Xu}, {Snyder}, {Nelson}, \&
  {Hernquist}}]{Vogelsberger14}
{Vogelsberger}, M., {Genel}, S., {Springel}, V., {et~al.} 2014, \mnras, 444,
  1518, \dodoi{10.1093/mnras/stu1536}

\bibitem[{{Wake} {et~al.}(2012){Wake}, {van Dokkum}, \& {Franx}}]{Wake12}
{Wake}, D.~A., {van Dokkum}, P.~G., \& {Franx}, M. 2012, \apjl, 751, L44,
  \dodoi{10.1088/2041-8205/751/2/L44}

\bibitem[{{Williams} {et~al.}(2009){Williams}, {Quadri}, {Franx}, {van Dokkum},
  \& {Labb{\'e}}}]{Williams09}
{Williams}, R.~J., {Quadri}, R.~F., {Franx}, M., {van Dokkum}, P., \&
  {Labb{\'e}}, I. 2009, \apj, 691, 1879, \dodoi{10.1088/0004-637X/691/2/1879}

\bibitem[{{Zahid} \& {Geller}(2017)}]{Zahid17}
{Zahid}, H.~J., \& {Geller}, M.~J. 2017, \apj, 841, 32,
  \dodoi{10.3847/1538-4357/aa7056}

\bibitem[{{Zahid} {et~al.}(2016){Zahid}, {Geller}, {Fabricant}, \&
  {Hwang}}]{Zahid16}
{Zahid}, H.~J., {Geller}, M.~J., {Fabricant}, D.~G., \& {Hwang}, H.~S. 2016,
  \apj, 832, 203, \dodoi{10.3847/0004-637X/832/2/203}

\bibitem[{{Zahid} {et~al.}(2018){Zahid}, {Sohn}, \& {Geller}}]{Zahid18}
{Zahid}, H.~J., {Sohn}, J., \& {Geller}, M.~J. 2018, \apj, 859, 96,
  \dodoi{10.3847/1538-4357/aabe31}

\end{thebibliography}
\bibliographystyle{aasjournal}

\end{document}